\documentclass[english,12pt]{article}
          \usepackage{babel}
          \usepackage[T1]{fontenc}
          \usepackage[latin2]{inputenc}
          \usepackage{pslatex}
\begin{document}
\begin{flushright}
\textit{ACTA ASTRONOMICA}\\
Vol.0 (2014) pp.0-00
\end{flushright}
\bigskip

\begin{center}
{\bf Hot spot manifestation in eclipsing dwarf nova HT Cassiopeiae}\\
\vspace{0.8cm}
{K. ~B~\k{a}~k~o~w~s~k~a$^{1}$ \& A. ~O~l~e~c~h$^1$}\\
\vspace{0.5cm}
\begin{small}
{$^1$ Nicolaus Copernicus Astronomical Center,
Polish Academy of Sciences,\\
ul.~Bartycka~18, 00-716~Warszawa, Poland\\}
\end{small}
{\tt e-mail: bakowska@camk.edu.pl}\\
~\\
\end{center}

\begin{abstract}  
We report the detection of the hot spot in light curves of the eclipsing dwarf nova HT Cassiopeiae during its superoutburst in 2010 November. Analysis of eight reconstructed light curves of the hot spot eclipses showed directly that the brightness of the hot spot was changing significantly during the superoutburst. Thereby, detected hot spot manifestation in HT Cas is the newest observational evidence for the EMT model for dwarf novae.  


\end{abstract}

\section{Introduction}

Among close binary systems there is a group called cataclysmic variables (CVs). CVs contain a white dwarf (the primary) and usually a main-sequence star (the secondary or the donor). In these kinds of systems, the donor fills its Roche-lobe and the material flows through the inner Lagrangian point from the secondary to the primary. In non-magnetic systems the matter spirals around the white dwarf forming a luminous accretion disk. The location where the material hits the edge of a disk is known as the hot spot or the bright spot. 

One of the subclasses of CVs are dwarf novae (DNs). In the light curves of DNs one can see regular outbursts with typical amplitudes of $2-5$ magnitudes. The interval between outbursts is from days (for the most active DNs like ER UMa stars) to tens of years. Among DNs stars there is a group named SU UMa type stars which are characterized by short orbital period ($P_{orb} < 2.5$ h). In the light curves of SU UMa stars not only outbursts are observed but also superoutbursts. Superoutbursts last longer than normal outbursts and are about one magnitude brighter. During superoutbursts one can see periodic light oscillations called superhumps in the light curves. A characteristic feature of SU UMa type stars is that the superhump period is a few percent longer than the orbital period  (more in Hellier 2001).

The eclipsing CVs (about $\sim 20\%$ of the total known, Warner 1995) provide a lot of valuable information (i.e. precise timing in the orbital motion or contribution of different components to the total light of the system). Worth noting is the fact that the analysis of the eclipse profile gives the successive stages of immersion and emersion of the primary and the hot spot. This information combined with simple geometric considerations allows us to obtain the relative sizes of the primary, the accretion disk, the hot spot and the secondary (Warner 1995). Due to these facts eclipsing DNs in particular during rare superoutbursts require further scrutiny.

HT Cassiopeiae was discovered over seventy years ago by Hoffmeister (1943) and classified as a U Gem type star with brightness varying between 13.0 and 16.5 mag. For 35 years this dwarf nova received very little attention, until the eclipses of HT Cas were noticed for the first time  (discovery made by Bond in 1978, private communication with Patterson). After this, Patterson made HT Cas a top priority object for an observing season in 1978. Three years later, the results of this campaign were published (Patterson 1981) where the first precise ephemeris of eclipses, ingress/egress times, inclination and mass ratio were presented. Additionally, Patterson characterized HT Cas as "the Rosetta stone of dwarf novae". Over 30 years from this statement, the literature about this unique eclipsing binary star is still growing, reaching several dozens of publications and several PhD theses.
 
Last year, Smak (2013b) brought attention to three eclipsing dwarf novae - HT Cas, NZ Boo and PU UMa and gave his interpretation of superhumps as being due to irradiation modulated periodically variable mass transfer rate. This shows that light curves of eclipsing dwarf novae are not only a good diagnostic tool for system parameters but also for theoretical interpretation of superoutburst and superhump mechanisms.

Nowadays, there is a hot discussion about constant or strongly enhanced mass transfer rate during superoutbursts in DNs (check Smak 2013a) it is a matter of the greatest importance to examine every possible eclipsing DN during superoutburst. The tool which provides information about transfer rate in CVs was invented by Smak (1994a). He presented the decomposition method which explains how to separate light from the disk and from the hot spot. Moreover, beautiful examples of the application of this method for DNs stars Z Cha and OY Car were presented by Smak (2007, 2008).

The newest example of an application of the decomposition method was given by Rutkowski et al. (2013) where an eclipsing CV denoted as HBHA 4705-03 was analysed. In the light curve of HBHA 4705-03, the hot spot was easily observed with a very high amplitude of $A_{s,max}=38.2\%$. In this case the analysis of the hot spot allowed the system parameters i.e. mean disk radius to be obtained.

In November 2010 HT Cas went into superoutburst (check Kato et al. 2012, B\k{a}kowska et al. 2014), which happened a quarter of a century after the previous, poorly observed, superoutburst. All these recent events were our motivation to make a detailed analysis of the hot spot and to verify the theory of superhump and superoutburst mechanisms.

This paper is constructed in the following way: section 2 describes observations of HT Cas during the superoutburst in 2010. In section 3 we present light curves of HT Cas. Later, in section 4 the results of the decomposition analysis are presented. Section 5 is a discussion chapter, where we analyse the presence of bright spots in CVs and we present the hot spot manifestation in HT Cas during the 1979-2010 observational seasons. The last section contains our conclusions.

\bigskip

\section{Observations and data reduction} 

To refer to our observations from now on we use only a day number HJD-2455000 [d]. Details about our observational data gathered during November 2010 HT Cas superoutburst were presented in B\k{a}kowska et al. (2014). Additonally, in Tab.1 we present our observation log covering HJD 503-513. During this part of the observational campagin three observers were gathering observations in Spain and USA, and four, small telescopes with diameters below 30 cm were used simultaneously. Moreover, data covering HJD 503-509 collected by the AAVSO\footnote{American Association of Variable Star Observers, www.aavso.org.} organization were used for this analysis. The exposure time and accuracy of photometric measurements depended on weather conditions, instruments used for observations and brightness of the object. The median value of photometric errors was 0.014 mag.

\vspace*{10.0cm}
\includegraphics{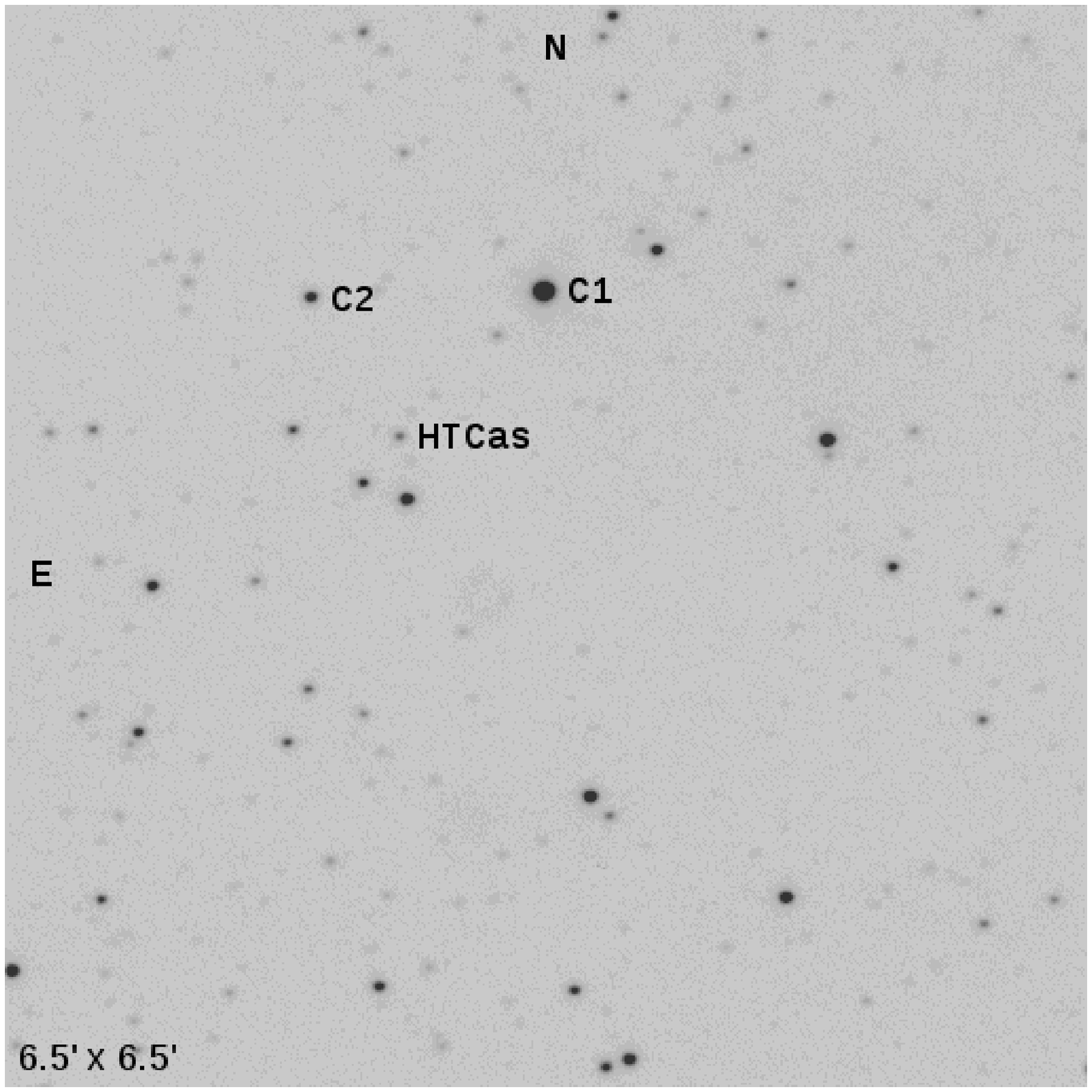}

   \begin{figure}[!ht]
      \caption {Finding chart of HT Cas. The field of view is about 6.5'$\times$6.5'. 
Both comparison stars are marked as C1 and C2, respectively. 
North is up, east is left.}
   \end{figure}

\begin{table*}
 \centering
 \begin{small}

  \caption{Journal of our CCD observations of HT Cas.}
  \begin{tabular}{@{}ccll@{}}
  \hline
   Observatory (Country)   & Telescope   & Observer & Date of observations    \\
                           &        &          &  HJD-2455000 [d]        \\
\hline
Antelope Hills Observatory (USA) & 10" &  R. Koff          & 503 - 508, 513\\
Observatorio del CIECEM  (Spain) & 0.10 m, 0.25 m &	E. de Miguel     & 503 - 506, 508 \\
TZEC Maun Foundation Observatory (USA) & 18 cm 	&  K. B\k{a}kowska  & 503, 505, 507, 513 \\
\hline
\end{tabular}
 \end{small}
\label{tab0}
\end{table*}

The relative unfiltered magnitude of HT Cas was obtained by taking the difference between the magnitude of the object and the mean magnitude of two comparison stars marked as C1 and C2 on Fig.1, where the map of the observed region is presented. The equatorial coordinates and brightness of comparison star C1 (RA=$01^{h}10^{m}06^{s}.709$, Dec=$+60^{o}05'25".31$, $11.92$ mag in $V_T$ filter) are taken from Tycho-2 Catalogue (Hog et al. 2000).

HT Cas was monitored in a clear filter ("white" light). Bias, dark current and flat-field corrections were made using the IRAF\footnote{IRAF is distributed by
the National Optical Astronomy Observatory, which is operated by the
Association of Universities for Research in Astronomy, Inc., under a
cooperative agreement with the National Science Foundation.} package. Profile photometry was obtained with DAOPHOTII (Stetson 1987). Relative manitudes were transformed to the standard Johnson V manitudes using data published by Henden \& Honeycutt (1997).

\section{Light curves}

In Fig.2 one can see the global photometric behaviour of HT Cas between HJD 503 (top panel) and HJD 513 (bottom panel). 

\vspace*{15.0cm}
\includegraphics{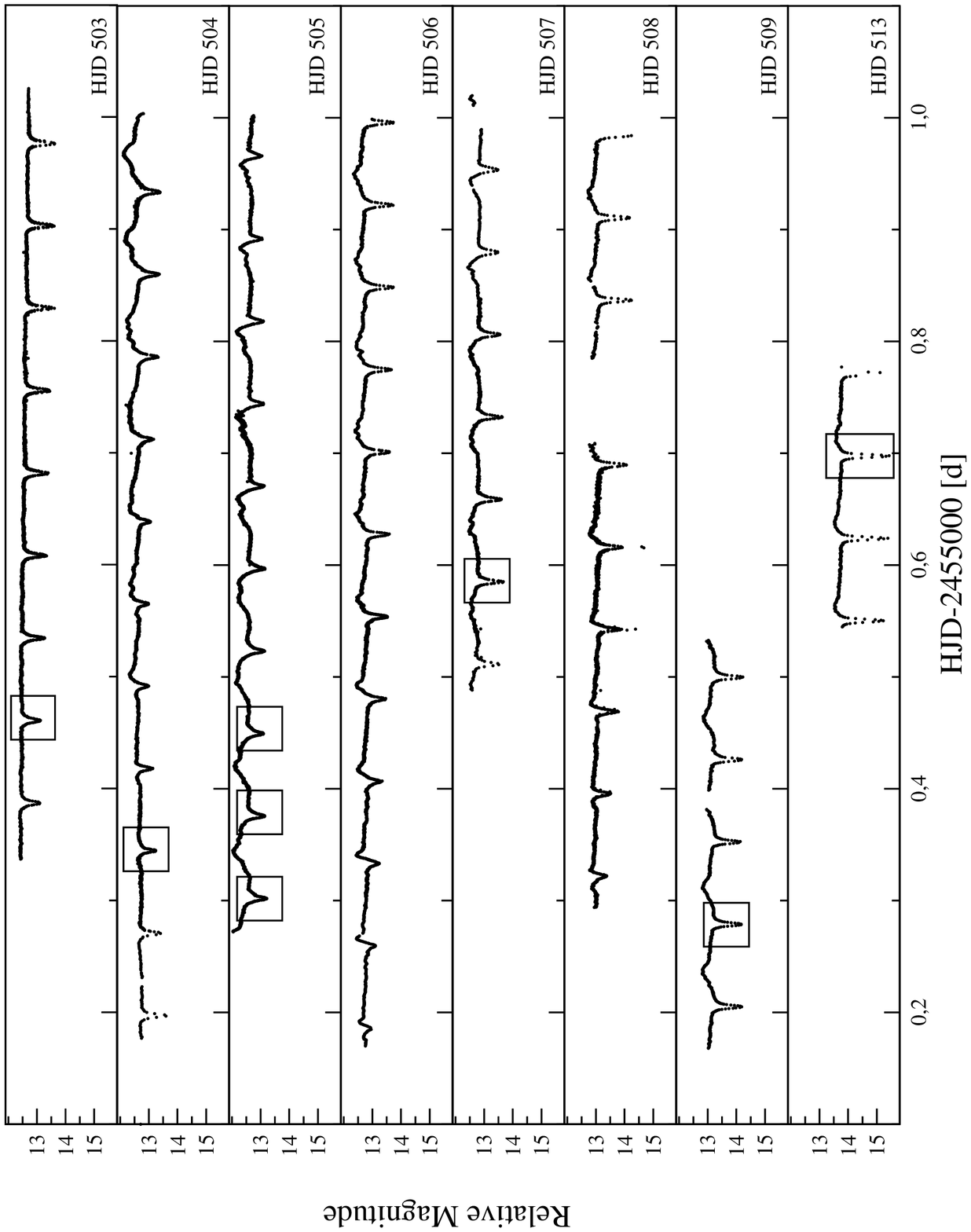}

   \begin{figure}[!ht]
      \caption {The global photometric behaviour of HT Cas during its November 2010 superoutburst. The first panel (top) presents the last night of the outburst precursor. The next six panels show six subsequent nights of the  superoutburst. In the bottom panel the light curve from the last night of the superoutburst is presented. The eclipses chosen for the decomposition method are marked by black, open squares. On $x-axis$ fraction of HJD is presented. Moreover, HJD-2455000 [d] is given on the right side of each panel.}
   \end{figure}

\vspace*{12.0cm}
\includegraphics{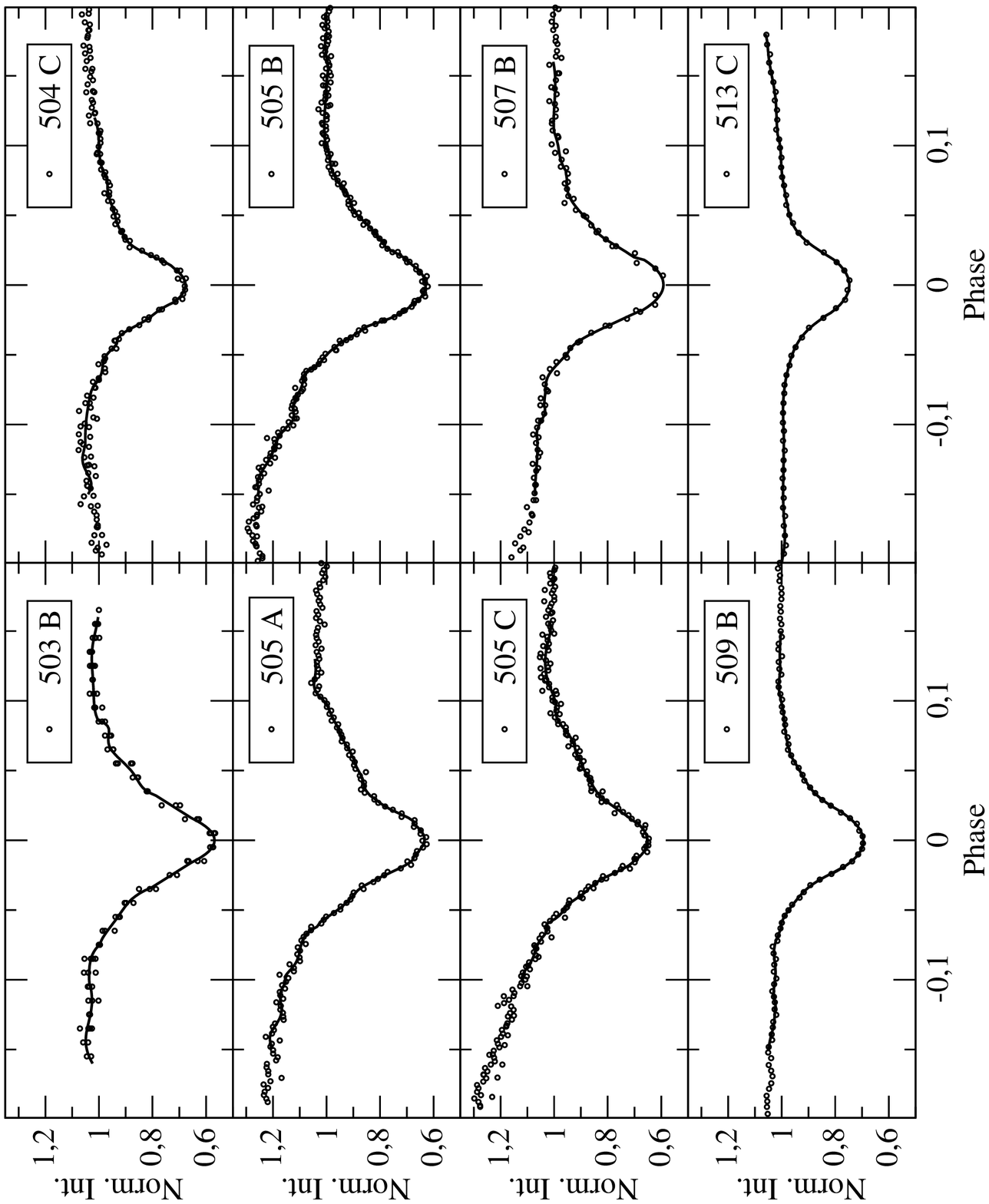}

   \begin{figure}[!ht]
      \caption {The eight eclipses chosen for decomposition analysis are presented. Observational data are displayed as open circles. Black lines represent synthetic light curves derived from the higher order polynomial fits. Those synthetic light curves were used for further analysis where separate light curves from the hot spot and the disk were obtained.}
   \end{figure}

We picked eight eclipses, marked in Fig.2 by black open rectangles for the decomposition analysis and they are presented in magnification in Fig.3. First eclipse 503 B was taken from the last night of the outburst precursor and allowed us to check the hot spot behaviour before the superoutburst. 
Next eclipse 504 C was the last eclipse just before the moment where the superhumps manifested their presence in fully developed form. Another three eclipses 505 A, 505 B and 505 C were the only eclipses without superhumps interfering in HJD 505 night. In HJD 506 night the superhumps maxima coincided with the eclipses and it was not possible to conduct a decomposition analysis. On the subsequent night, we were fortunate and from the light curve HJD 507 we could choose easily an eclipse 507 B for further calculations. In HJD 508 one more time, the superhumps excluded the eclipses from further analysis. Again, the situation changed significantly on the next night HJD 509, from which we obtained good results and for the decomposition analysis the second eclipse, denoted as 509 B, was chosen. Although we gathered some observational material from HJD 510-512, those data were not used for the hot spot analysis. Observations from HJD 510 and HJD 512 due to weather condition were scattered and with higher level of noise than the rest of our data. During HJD 511 superhump maxima coincided with eclipses and we could not pick any eclipse from our observational runs. The HJD 513 night was the last night of the superoutburst. The 513 C eclipse was the last possible eclipse to choose for the decomposition analysis from our observational set of data. 

\section{Results}

The decomposition method invented by Smak (1994a) allowed us to obtain eight separate light curves of the hot spot and the disk. We fitted the uneclipsed parts of the resulting hot spot light curves with a formula given by Paczy\'{n}ski \& Schwarzenberg-Czerny (1980)
\begin{equation}
l^*_{s,o}(\phi)=A_{s,max}[1-u+u\cos(\phi-\phi_{max})]\cos(\phi-\phi_{max})
\label{eq3}
\end{equation} 
with the limb darkening coefficient $u=0.6$ and in Tab.2 we display the hot spot amplitude $A_{s,max}$, phase of its maximum $\phi_{max}$ and phases of contacts $\phi_1$, $\phi_2$, $\phi_3$ and $\phi_4$. In Fig.4. (from top panel to bottom) the spot light curves were displayed from the beginning untill the end of the superoutburst. The hot spot had a rather small amplitude of brightness during the early stages of the superoutburst (503 B and 504 C), where the hump amplitude was below $5\%$. Later, when the superoutburst hit maximum brightness (505 A, 505 B and 505 C), the hot spot was presenting itself in a fully developed form with the amplitude over $10\%$. During the plateau stage of the HT Cas superoutburst (507 B and 509 B), the "standard" hot spot was present with the average amplitude between $3 - 8\%$. The most surprising result was found during last night of the superoutburst (513C), when the hot spot was absent. In Fig.5 we display the light curve of the superoutburst (top panel) and the amplitude of the hot spot (bottom panel). In our opinion, there is a strong dependence between the phase of the superoutburst and the hot spot brightness. This fact implies that the mass transfer during superoutburst in strongly enhanced and we discuss this fact in the following Section. 

\begin{table*}
\begin{small}
 \begin{center}
  \caption{Hot Spot in HT Cas. Hump parameters and phase contacts.}
  \begin{tabular}{@{}lllllll@{}}
  \hline
   No. of     &  Phase of hot spot   & Hot spot  & Phases of &  &  & \\
  eclipse  					 &  max. brightness      &         amplitude           & contact & & & \\
   & $\phi_{max}$ & $A_{s,max}$[\%]& $\phi_1$ & $\phi_2$ & $\phi_3$ & $\phi_4$ \\
\hline
503 B & $-0.080\pm0.009$ & $3.96\pm0.14$ & $-0.080\pm0.01$ & $-0.020\pm0.01$ & $0.085\pm0.01$ & $0.105\pm0.01$\\ 
504 C & $-0.087\pm0.016$ & $4.73\pm0.18$ & $-0.085\pm0.01$ & $-0.020\pm0.01$ & $0.105\pm0.01$ & $0.120\pm0.01$\\ 
505 A & $-0.073\pm0.002$ & $14.70\pm0.15$ & $-0.065\pm0.01$ & $-0.015\pm0.015$ & $0.100\pm0.005$ & $0.115\pm0.005$\\ 
505 B & $-0.086\pm0.003$ & $13.96\pm0.21$ & $-0.087\pm0.005$ & $-0.023\pm0.003$ & $0.090\pm0.005$ & $0.115\pm0.01$\\ 
505 C & $-0.066\pm0.004$ & $10.88\pm0.27$ & $-0.060\pm0.005$ & $-0.015\pm0.005$ & $0.100\pm0.01$ & $0.120\pm0.01$\\ 
507 B & $-0.089\pm0.003$ & $7.74\pm0.08$ & $-0.080\pm0.005$ & $-0.020\pm0.003$ & $0.100\pm0.005$ & $0.120\pm0.005$\\ 
509 B & $-0.052\pm0.003$ & $3.85\pm0.05$ & $-0.055\pm0.005$ & $-0.025\pm0.005$ & $0.075\pm0.005$ & $0.105\pm0.005$\\ 
513 C & - & $0.36\pm0.07$ & - & - & - & -\\ 
\hline
\end{tabular}
\label{tab3}
\end{center}
\end{small}
\end{table*}

\section{Discussion}

\subsection{Summary of the theory}

The debate about eruption mechanisms in DNs is long-lasting, very hot and intense (see Osaki \& Kato 2013a, Smak 2013a). 

According to Osaki (1974), the outbursts of DNs are caused by sudden gravitational energy release due to accretion of matter onto the primary component and a steady mass loss from the secondary is assumed. At the same time, Bath (1971) proposed a similar model for the outbursts, but with one crucial difference, the outbursts are assumed to be due to the unsteady mass overflow.

\newpage

\vspace*{22.0cm}
\includegraphics{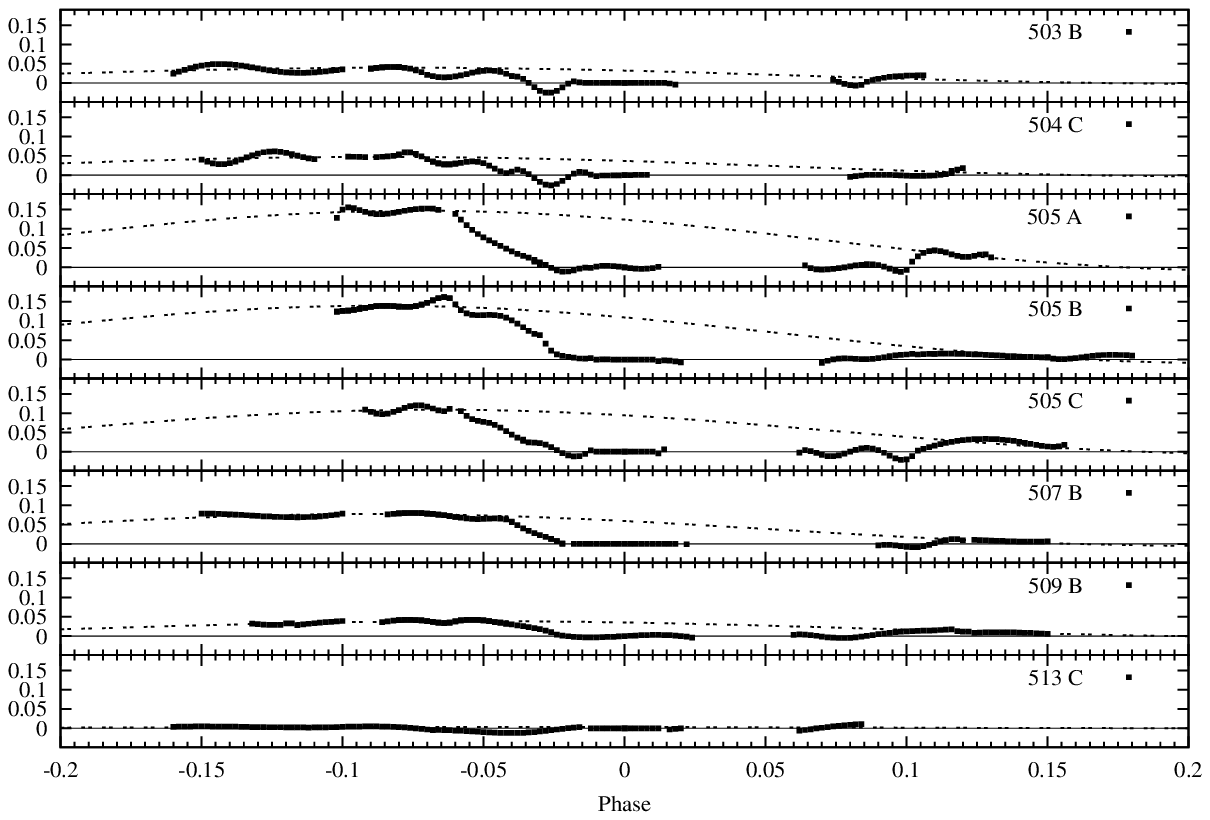}
\vspace*{-2.0cm}
   \begin{figure}[!ht]
      \caption {Reconstructed hot spot light curves of HT Cas during its November 2010 superoutburst. Dotted lines represent Eq.(1) with parameters listed in Tab.2. The hot spot manifestation is clearly seen, especially on panels 3-5 (from top), where the brightness amplitude of the hot spot was higher than $10\%$. During the last night of the superoutburst, the hot spot was not observed (bottom panel) and its amplitude was below $0.5\%$.}
   \end{figure}

\vspace*{12.0cm}
\includegraphics{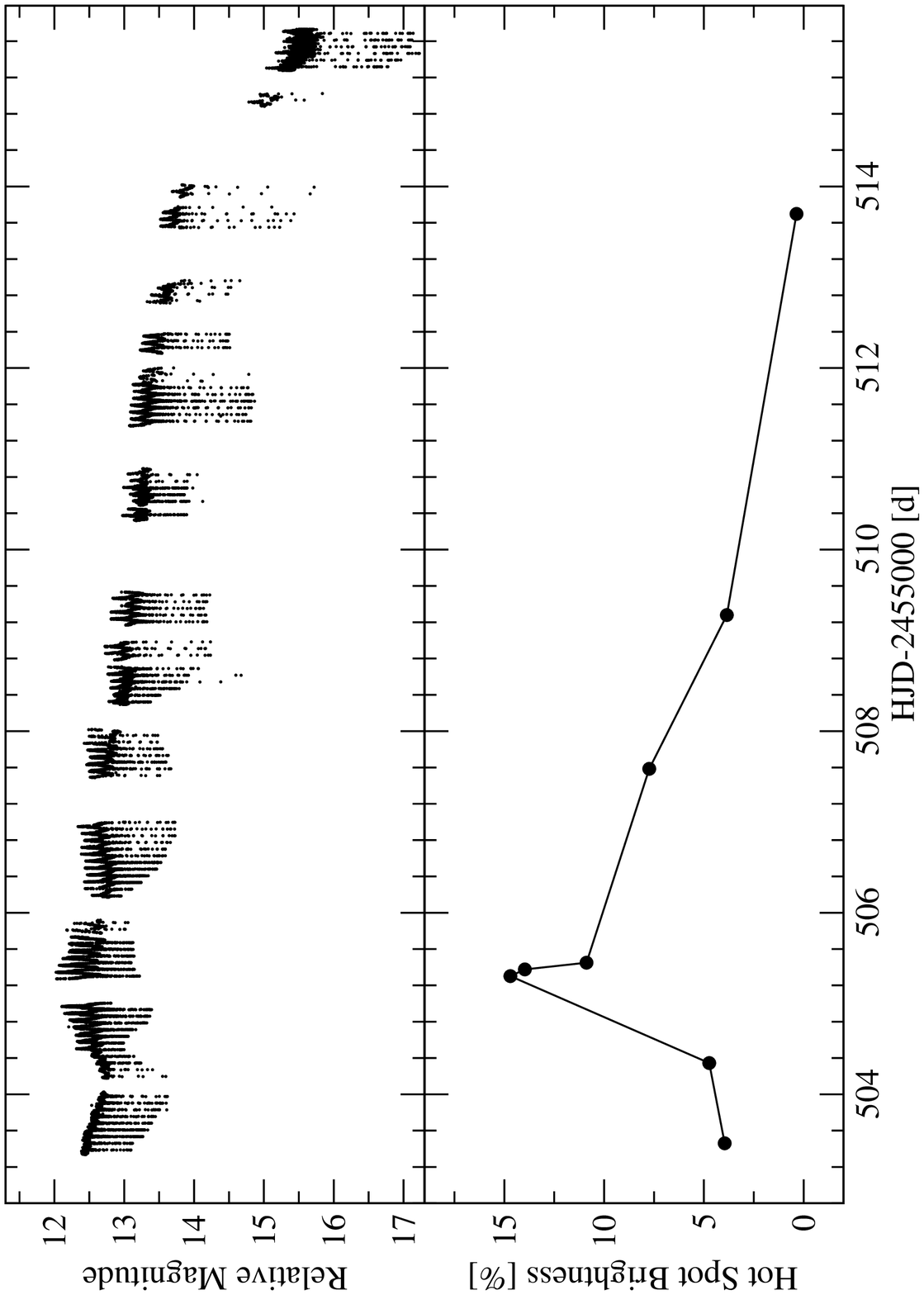}

   \begin{figure}[!ht]
      \caption {In the top panel we present consecutive nights of the HT Cas superoutburst in November 2010. Our observations started during the outburst precursor (on HJD 503) and ended when HT Cas reached quiescence level (on HJD 518). On the bottom panel we display the amplitude of the hot spot during the superoutburst. There is a strong correlation between the spot amplitude and the stage of the superoutburst. At the beginning of the superoutburst the hot spot amplitude was below $5\%$. When the superoutburst reached maximum brightness, the hot spot amplitude also was the highest ($A_{s,max}>10\%$). During the last night of the superoutburst (HJD 513) the hot spot was not detected ($A_{s,max}<0.5\%$).}
   \end{figure}

In the meantime, Osaki (1989) proposed the Thermal-Tidal Instability (TTI) model in which thermal instability is coupled with the tidal instability. Normal outbursts are explained in terms of the thermal limit-cycle instability in the accretion disk (see Cannizzo 1993, Lasota 2001). The superhumps mechanism is understood by considering the tidal instability (Whitehurst 1988, Hirose \& Osaki 1990, Lubow 1991). The TTI model is based on constant mass-transfer rate from the secondary and all variability is thought to be produced within the disk (Osaki 1996, Osaki 2005 for a review).  

Nevertheless, there was a serious criticism concerning the TTI model. Vogt (1983) observed variation of the orbital hump amplitude in VW Hyd and he postulated the enhanced mass transfer from the secondary star (the EMT model). Schreiber et al. (2004) compared the TTI and the EMT models, and found that for the observed variations in the optical light curve of VW Hyi the EMT model is more promising scenario. As a solution for the superoutburst and superhump mechanisms Smak (2004a, 2004b) proposed a hybrid model which combines the TTI model with the irradiation-enhanced mass-transfer effects. 

Moreover, Buat-M\'{e}nard \& Hameury (2002) presented the mass accretion rate from the secondary (Fig.5 top panel, dashed line) using the parameters of EG Cnc. They showed that the accretion rate rises in the beginning of the superoutburst. Later, during maximum brightness of the superoutburst, there is a plateau phase of the mass accretion rate. At last, when the star comes back to the quiescence level, also the mass accretion rate falls down. Based on the estimations of Buat-M\'{e}nard \& Hameury (2002), one can see a correlation between the changes of the mass accretion rate from the secondary and the stage of the superoutburst. Additionally, Smak (1993) calculated the mass-transfer rate at quiescence for WZ Sge, estimated from the luminosity of the hot spot (Eq.10). There is a rough linear dependence between the luminosity of the hot spot and the mass-transfer rate. All these facts are not only in favour of the EMT model, but also confirm our results displayed in Fig.5. 

In a recent discussion concerning superoutburst and superhump mechanisms (check Osaki \& Kato 2013a, Smak 2013a) changes of the hot spot manifestation during superoutburst in HT Cas were atributed to the enhanced mass transfer rate. We believe that data collected during
November 2010 superoutburst of HT Cas further support this thesis.

\begin{table*}
\begin{small}
 \begin{center}
  \caption{Hot spot manifestation in cataclysmic variables.}
  \begin{tabular}{@{}llllll@{}}
  \hline
   Object &  Date   & Hot spot amplitude & Type of object & Star activity & Author  \\
   	    	  & [year]  & [$\%$]             &                & \\
  
\hline
UX UMa 	& 1952, 1953, 1955 & $20.1\% - 13.5\%$ & nova-like & quiescence  & Smak (1994b) \\
		& 1958, 1961, 1974 & $7.4\%-4.9\%$	  & 			  &			   & Smak (1994c)\\
RW Tri 	& 1994			   &	 $14.1\%-15.7\%$  & nova-like & quiescence  & Smak (1995)\\  
 Z Cha 	& 1973, 1980-82,  & $17.7\% - 4.3\%$ & dwarf nova & superoutburst  & Smak (2007)\\ 
		& 1984 & $8.0\%-0\%$      &            & superoutburst          &             \\
OY Car 	& 1980 & $17.6\% - 13.4\%$ & dwarf nova & superoutburst  & Smak (2008) \\ 
		& 1984 & - & quiescence & \\
HBHA 4705-03 & 2010 & $38.2\%$ & nova-like & quiescence & Rutkowski et al. (2013) \\ 
\hline
\end{tabular}
\label{tab4}
\end{center}
\end{small}
\end{table*}

\begin{table*}
 \begin{center}6
  \caption{Visual inspection of the hot spot manifestation in HT Cas in 1978-2010.}
  \begin{tabular}{@{}lllll@{}}
  \hline
   Date   &    No. of eclipses & Hot spot  & Star activity & Author  \\
		  &                    & manifestation                       &               &         \\   
   
  \hline
 1978 Aug-Dec  &        8      & present          &  quiescence  &  Patterson (1981)\\
 1978 Dec	   & 		4	  &	weak or absent 		    & outburst	& 					 \\
 1983 Nov-Dec  &     	10    & weak or absent 	&  quiescence  &  Zhang et al. (1986)\\
 1985 Jan 	   &			4	  &	present 	   		& superoutburst     &					\\
 1982 Sep      &     	4     & weak or absent   & quiescence &  Horne et al. (1991)\\
 1983 Nov-Dec  & 		12    & weak or absent   & quiescence &                    \\ 
 1995 Nov      &   		7     &  absent          & outburst  &  Ioannou et al. (1999)\\
 1997 Mar      &			1	  &	 absent 		   	 &	 quiescence	&						\\
 2002 Sep      & 		2     & present          & 'low' quiescence &  Feline et al. (2005)\\ 
 2003 Oct      & 		3     &	weak or absent   &	 'high' quiescence &						\\
 2007-2009     &        63	  & weak or absent   & 'low' and 'high' quiescence & Baptista et al. (2011)\\
 2010 Nov      &         8      & present         & superoutburst &  this work\\ 
\hline
\end{tabular}
\label{tab5}
\end{center}
\end{table*}

\subsection{Hot spot manifestation in CVs}

Until now, the decomposition method for hot spot manifestation was used by Smak (1994b, 1994c, 1995, 2007, 2008) and Rutkowski et al. (2013) for five eclipsing CVs. In Tab.3 we present a set of basic statistics based on their results for nova-like (NLs) objects UX UMa, RW Tri and HBHA 4705-03 and for DNs stars Z Cha and OY Car. 

NLs are recognized as resembling novae between eruptions Warner (1995) and the key difference between NLs and DNs is defined by the rate at which the mass is supplied to the disk (see Knigge et al. 2011 for a review). Based on the decomposition analysis conducted for DNs, it is possible to determine mass transfer rate during superoutbursts. Due to this fact, we now look carefully for bright spot analysis made for objects OY Car and Z Cha. Two eclipses of OY Car observed during its November 1980 superoutburst were analysed by Smak (2008) based on observations made by Schoembs (1986). The hot spot manifested its presence with high amplitudes $A_{s,max}=17.6(9)\%$ and $A_{s,max}=13.4(3)\%$. For comparison, Smak (2008) analysed OY Car during quiescence in 1984 using observations made by Wood et al. (1989). This time the hot spot was absent. The conclusion given by Smak (2008) was that the mass transfer rate in OY Car during its superoutbursts is strongly enhanced. 

Now, let's look closer at some of the system parameters for Z Cha and HT Cas, respectively, i.e. inclinations $i = 81.8^o$ (Ritter \& Kolb 2003) and $i=81^o$ (Horne et al. 1991), and also orbital periods $P_{orb}=0.0744994$ days (Mumford 1971) and  $P_{orb}=0.0736469$ days (B\k{a}kowska et al. 2014). The immediate conclusion is that Z Cha is very similar to HT Cas. Smak (2007) examined the eclipse profiles of Z Cha during its six superoutbursts between January 1973 and the December 1984 based on observations made by Warner \& O'Donoghue (1988). From decomposition analysis, the variety of the bright spot amplitudes was derived, from no detection during December 1984 superoutburst (Smak 2007, Fig.2) to the highest amplitude of $A_{s,max}=17.7\%$. A thorough examination of hot spot manifestation in Z Cha led to the same conclusions, as in the case of OY Car, that the mass transfer rate during superoutbursts is strongly enhanced and that the tidal-thermal instability model under the assumption of constant mass transfer rate should be seriously modified or even abandoned. These concluding remarks were questioned by Osaki \& Kato (2013b). They claim that observational data of Z Cha, used for decomposition analysis, did not cover the full duration (10-12 days) of the plateau phase of any superoutburst and they are not statistically significant enough to test the TTI and the EMT models. This remark is no longer up-to-date due to the well observed HT Cas November 2010 superoutburst. During this unique phenomenon, we analysed eclipse profiles from the outburst precursor till the last night of the superoutburst. Thereby, we have the same conclusion as Smak (2007, 2008), that the hot spot manifestation confirms strongly enhanced mass transfer during superoutbursts. 

\subsection{Our observations in context}

In Tab.4 we present a visual inspection of hot spot manifestation in HT Cas based on observations made in 1978-2010. We examined the HT Cas light curves presented by Patterson (1981), Zhang et al. (1986), Horne et al. (1991), Ioannou et al. (1999), Feline et al. (2005) and Baptista et al. (2011). We could only report whether the hot spot was present or absent, any other more detailed conclusions were not possible. Nevertheless, a rough analysis of the HT Cas light curves, collected during the quiescent state, showed also the variety of the hot spot amplitudes. Does it mean that the mass transfer rate is unstable also between outbursts and superoutbursts? In our opinion, it would be great to use the decomposition method for data from quiescence and see exact results.

\section{Conclusions}

We report the detection of the hot spot in eclipsing dwarf nova HT Cas during its November 2010 superoutburst. To obtain separate light curves from the disk and from the hot spot, we used the decomposition method presented by Smak (1994a). Our results show that the hot spot amplitude was changing rapidly during superoutburst. Moreover, the brightness of the bright spot was correlated with the stage of the superoutburst. During the outburst precursor and the first night of the superoutburst the hot spot amplitude was below $A<5\%$. When the superoutburst reached its maximum brightness, also the hot spot had the highest amplitude with a value around $A\sim 14\%$. At the last stages of the superoutburst, just before returning to the minimum brightness, the hot spot was not detected.  Taking into account these facts, the hot spot manifestation in November 2010 superoutburst in HT Cas is the newest observational evidence for the enhanced-mass transfer rate postulated by the EMT model proposed by Smak (2013b).

\bigskip 

\noindent {\bf Acknowledgments.} ~We want to thank Prof. Smak for his thoughtful comments about the decomposition method and Dr. Alexis Smith for language corrections. We acknowledge generous allocation of the Tzec Moun Foundation for telescope time.  Data from AAVSO observers are also appreciated. Project was supported by  Polish National Science Center grant awarded by decision DEC-2012/07/N/ST9/04172 for KB.

\end{document}